# Observation of robust intrinsic C points generation with magneto-optical bound states in the continuum

Wenjing Lv[1,2,#], Haoye Qin[3,#], Zengping Su[1,2], Chengzhi Zhang[1], Jiongpeng Huang[1,2], Yuzhi Shi[4], Bo Li[1,5], Patrice Genevet[6], Qinghua Song[1,5,*]

[1]*Tsinghua Shenzhen International Graduate School, Tsinghua University, Shenzhen, 518055, China*

[2]*School of Materials Science and Engineering, Tsinghua University, Beijing, 100084, China*

[3]*Laboratory of Wave Engineering, École Polytechnique Fédérale de Lausanne, 1015 Lausanne, Switzerland*

[4]*Institute of Precision Optical Engineering, School of Physics Science and Engineering, Tongji University, Shanghai 200092, China*

[5]*Suzhou Laboratory, Suzhou 215000, China*

[6]*Physics Department, Colorado School of Mines, 1523 Illinois St., Golden, CO 80401, USA*

[#]*These authors contributed equally: Wenjing Lv, Haoye Qin*

*Corresponding author. Email: song.qinghua@sz.tsinghua.edu.cn*

**Abstract**

C points, characterized by circular polarization in momentum space, play crucial roles in chiral wave manipulations. However, conventional approaches of achieving intrinsic C points using photonic crystals with broken symmetries suffer from low Q factor and are highly sensitive to structural geometry, rendering them fragile and susceptible to perturbations and disorders. In this letter, we report the realization of magneto-optical (MO) bound states in the continuum (BICs) using a symmetry-preserved planar photonic crystal, achieving intrinsic at-$\Gamma$ C points that are robust against variation in structural geometry and external magnetic field. MO coupling between two dipole modes induces Zeeman splitting of the eigenfrequencies, leading to MO BICs and quasi-BICs with circular eigenstates for high-Q chiral responses. Furthermore, switchable C point handedness and circular dichroism are enabled by reversing the magnetic field. These findings unveil a new type of BICs with circular eigenstates and on-demand control of C points, paving the way for advanced chiral wave manipulation with enhanced light-matter interaction.

Chiro-optical effect, characterized by distinct responses to right- and left-circularly polarized light, remains pivotal across diverse applications, encompassing chiral light sources, chiral detectors, chiral sensing techniques, valleytronics, and asymmetric photocatalysis, among others [1–5]. This phenomenon typically arises from extrinsic or intrinsic chirality, associated respectively with oblique excitation (off-Γ point) and normal excitation (Γ point) [1,5–7]. Specifically, the chiro-optical effect induces a transition of the polarization field in momentum space from linear to elliptical, or even to circular polarization (widely known as C point). The C point at-Γ is particularly esteemed and valued, especially in the realm of chiral emission and polarized photodetection [6,8]. To circumvent the fabrication challenges posed by 3D metamaterials [9–11], researchers have proposed achieving strong intrinsic chirality through resonant planar metasurfaces with corner-like geometry [1,12,13], and more recently, slant geometries that break both in-plane and out-of-plane symmetries, potentially yielding near-unity circular dichroism and chiral emission [6,14]. The quality factor (Q factor) of chiral metasurfaces is also critical in defining chiral light-matter interactions and chiro-optic responses in practical applications [15]. To enhance the Q factor, the concept of bound states in the continuum (BICs) [16,17] has been introduced to realize so-called intrinsic chiral BICs [6]. However, achieving this necessitates breaking inversion symmetry to shift the off-Γ C points toward Γ point [18,19], which inevitably compromises the Q factor of such devices [20]. Moreover, the sensitivity of these C points to geometric configurations renders them fragile and susceptible to perturbations and disorders.

Here, we propose the concept of magneto-optical (MO) BIC empowered by a magnetic PhC and demonstrate its capability in robustly generating intrinsic high-Q C points at Γ point. Our approach leverages MO coupling between two dipole modes in the PhC. Theoretical analysis confirms that by applying external magnetic field, the eigenstates induced by such coupling consistently maintain circular states, laying the foundation for constant C point generation. By adjusting the magnetic field, this mode coupling leads to the cancellation of radiation loss, thereby achieving an MO BIC with an infinite Q factor. Furthermore, to illustrate the versatility of our approach in manipulating chiral waves, we demonstrate high Q chiral emission and chiral transmission with near-unity circular dichroism.

The proposed PhC is composed of MO material featuring an array of squared air holes with length and thickness denoted as $L$ and $t$ as shown in Fig. 1a. The MO material can be described by a permittivity tensor with off-diagonal terms as [21–23],

$$\varepsilon = \begin{pmatrix} \varepsilon_0 & -i\varepsilon_z & 0 \\ i\varepsilon_z & \varepsilon_0 & 0 \\ 0 & 0 & \varepsilon_0 \end{pmatrix} \quad (1)$$

where $i^2 = -1$, $\varepsilon_0$ represents the zero-field dielectric permittivity, and $\varepsilon_z$ represents the MO coupling term proportional to the applied magnetic field $H_z$ along z-direction. Thus, by applying an external magnetic field, the MO coupling can be induced and controlled. Since the permittivity $\varepsilon_0$ solely influences the resonant frequency, we assume $\varepsilon_0 = 8$ without loss of generality in the following discussions. Initially, we disable the MO coupling by setting the magnetic field to zero, i.e., $\varepsilon_z = 0$. The eigenmode profiles of degenerated $|e_{1,2}\rangle$ exhibits 90° rotation symmetry, confirming them as orthogonal dipoles characterized by orthogonal linear polarizations due to the unbroken inversion symmetry as shown in Fig.1b. We find that the band structure exhibits a relatively low Q factor, with the far-field polarization state being linear (Fig. 1f), denoted as an L point (blue dot on Poincaré sphere in Fig. 1e). Upon activating the magnetic field ($\varepsilon_z \neq 0$), it's observed that the field distribution of the MO coupled modes (Fig. 1d) precisely matches the eigenmode profile obtained from the superposition of the previous two dipoles, i.e., $|e_1\rangle \pm i|e_2\rangle$ as shown in Fig.1c. This implies that the MO coupled modes $|\pm\rangle$ should exhibit eigenstates of $[1, \pm i]^T$ under $|e_{1,2}\rangle$ basis, manifesting as circular-polarization radiation. Therefore, when $\varepsilon_z \neq 0$, robust at-Γ C points can be obtained (red dots for $|-\rangle$ as shown in Fig. 1e). For example, when $\varepsilon_z = 0.04$, a high Q factor occurs around Γ point, implying the formation of MO quasi-BIC (QBIC) as shown in Fig. 1g (left). As expected, the corresponding polarization states of such high-Q band reveal C points at-Γ in the momentum space (right panel of Fig. 1g). At a critical value of $\varepsilon_z = 0.0625$ in Fig. 1h, an MO BIC is realized on $|-\rangle$ band with an infinite Q factor at Γ point, resulting in a non-radiative MO V point denoted as a black dot in Fig. 1e. Interestingly, we observe a reversal in the handedness of the surrounding polarization states near the MO BIC (see more details in Fig. S1). It's worth noting that such MO BIC occurs under unchanged robust circular eigenstates, i.e., C points, while manifesting as far-field V point due to the elimination of radiation, which is inaccessible in conventional configurations.

We proceed to delve into unraveling the underlying mechanism of MO coupling induced BIC and the formation of robust C points. The MO effect is conceptualized as the coupling between two orthogonal degenerated dipole modes $|e_{1,2}\rangle$ with the Hamiltonian given as,

$$H = \begin{pmatrix} \omega - i\gamma & V_{12} \\ -V_{12} & \omega - i\gamma \end{pmatrix} \quad (2)$$

where the two diagonal terms $\omega - i\gamma$ represent the complex frequency, which are equal to each other due to the $C_2$-rotation symmetry of the PhC structures, $V_{12} = \frac{i}{2} \frac{\sqrt{\omega_1 \omega_2} \int \varepsilon_z \hat{z}(E_1^* \times E_2) dV}{\sqrt{\int \varepsilon_0 |E_1|^2 dV \int \varepsilon_0 |E_2|^2 dV}}$ is the complex coupling strength between the two dipole modes [24,25], $E_{1,2}$ is the electric field. Therefore, the two complex eigenfrequencies $\omega_{1,2}^{MO}$ of the Hamiltonian are given as,

$$\omega_{1,2}^{MO} = \omega - i\gamma \pm iV_{12} \quad (3)$$

In the absence of magnetic field ($\varepsilon_z = 0$), the coupling strength $V_{12}$ vanishes, resulting in the degeneracy of two modes as a Dirac point ($\omega_{1,2}^{MO} = \omega - i\gamma$). When $\varepsilon_z \neq 0$, the complex coupling strength $V_{12}$ becomes non-zero and proportional to $\varepsilon_z$ (see Supplementary Fig. S2), leading to a splitting of both the real and imaginary part of the two complex eigenfrequencies as shown in Figs. 2a and 2b, respectively, which is also known as Zeeman splitting (see also Supplementary Fig. S3 for both bands with negative and positive $\varepsilon_z$). It can be seen that the extracted complex eigenfrequencies from simulation are in line with theoretical predictions. It should be noted that the imaginary part of the complex eigenfrequency is typically represented as the loss or Q factor. With the gradual increase of MO coupling strength, one eigenmode exhibits lower loss rate, approaching BIC conditions with zero imaginary part and infinite Q factor at a specific point denoted as $|-\rangle_{BIC}$, and then deviate as shown in Fig. 2b. For the other band, $|+\rangle_{BIC}$ is achievable with negative $\varepsilon_z$ and inverse circular eigenstates (see Supplementary Fig. S3 and S4). When $\varepsilon_z \neq 0$, the two eigenstates of the Hamiltonian in Eq. 2 can be derived as,

$$e_{1,2}^{MO} = [1, \pm i]^T \quad (4)$$

representing the anticipated circular states denoted as $|\pm\rangle$ in this paper. It is noteworthy that the Zeeman splitting induced by the coupling between the two dipole modes plays a crucial role in the circular eigenstates generation and the resulting intrinsic at-Γ C points, which cannot be achieved with uncoupled modes [26]. Consequently, the far-field polarization at Γ point consistently maintains circular polarization with Stokes parameter $S_3 = \pm 1$, except around exact BIC conditions or $\varepsilon_z = 0$, as shown in Fig. 2c. The corresponding polarization ellipses are plotted in Fig. 2d, verifying the generation of robust C points across varying $\varepsilon_z$ ranging from 0.01 to 0.14.

Subsequently, we evaluate the immunity of C points against parameter variations, specifically focusing on the side length (*L*) of squared air holes and the thickness (*t*) of the PhC slab as two evaluation parameters. The extracted far-field polarization Stokes parameter $S_3$ and Q factor of the two eigenmodes are shown in Fig. 3a. It's observed that the C point ($S_3 = \pm 1$) persists across all evaluated scenarios for both bands, except when approaching MO BIC conditions. This result holds true for all non-zero $\varepsilon_z$, including very small value of $\varepsilon_z$ as shown in Fig. 3b, further underscoring the robust generation of intrinsic C points with high Q factor.

To illustrate the potential applications of the proposed magnetic PhCs, we conducted additional evaluation of switchable chiral emission and stable circular dichroism (CD) as shown in Fig. 4. The intrinsic chiral emission spectra for specific values $\varepsilon_z = 0$, $\pm 0.02, \pm 0.04, \pm 0.06$ and $\pm 0.08$ are shown in Fig. 4a, revealing sharp linewidths and high contrast between LCP and RCP emissions. Near MO BIC wavelength, strong emission intensity occurs. Notably, the emission chirality (LCP or RCP) can be controlled by switching the sign of $\varepsilon_z$. The corresponding emission CD values remain near 1 and -1 for positive and negative $\varepsilon_z$, implying the potential for tunable MO BIC lasers. Furthermore, transmission and the associated CD are also demonstrated in Supplementary Fig. S5.

It should be noted that the MO BIC and robust intrinsic C points can be achieved in practice uusing low-loss MO thin-film materials ssuch as bismuth iron garnet (BIG) [21,24,27], bismuth doped Yttrium Iron Garnet (Bi:YIG) [28,29], and Cerium doped YIG (Ce:YIG) [22,30], followed by standard e-beam lithography and etching processes in nanofabrication of metasurfaces and PhCs.

In conclusion, we have presented the concept of MO BIC and the robust generation of intrinsic C points with high-Q factor using a magnetic planar PhC featured by simple C2 rotation symmetry. We discover that the MO BIC manifests as a V-point singularity within circular polarizations, a characteristic not commonly accessible in typical PhC. Thanks to MO coupling, these C points exhibit robustness against structural geometry variations with non-zero external magnetic field. The manipulation of C points handedness and circular dichroism is enabled by reversing the sign of magnetic field. The proposed mechanisms may contribute to strongly enhancement of chiral light-matter interaction, thereby improving chiral emission/detection efficiency, facilitating reconfigurability of chiral responses [31], enabling robust chiral meta-devices. Furthermore, the magnetic field introduces an additional dimension for controlling, realizing, and switching BIC and other singularities for practical applications in topology, optics and quantum [32–34].

*Note added*. —During the preparation of the manuscript, we became aware of Ref. [35].

**Acknowledgments**

Q.S. acknowledges the funding support from the National Key R&D Program of China (No. 2023YFB3811400), the National Natural Science Foundation of China (No. 12204264), the Shenzhen Science and Technology Innovation Commission (No. WDZC20220810152404001; JCYJ20230807111706014; JSGGZD20220822095603006). Y.S. acknowledges the National Key Research and Development Program of China (No. 2023YFF0613600), the Shanghai Pilot Program for Basic Research, National Natural Science Foundation of China (No. 62205246), and Science and Technology Commission of Shanghai Municipality (No. 22ZR1432400).

**Conflicts of interest**

The authors disclose no conflict of interest.

**References**

[1] A. Y. Zhu, W. T. Chen, A. Zaidi, Y.-W. Huang, M. Khorasaninejad, V. Sanjeev, C.-W. Qiu, and F. Capasso, *Giant Intrinsic Chiro-Optical Activity in Planar Dielectric Nanostructures*, Light Sci. Appl. **7**, 2 (2018).

[2] E. Plum, V. A. Fedotov, and N. I. Zheludev, *Extrinsic Electromagnetic Chirality in Metamaterials*, J. Opt. A: Pure Appl. Opt. **11**, 074009 (2009).

[3] Y. Chen et al., *Multidimensional Nanoscopic Chiroptics*, Nat Rev Phys **4**, 113 (2022).

[4] X. Duan, B. Wang, K. Rong, C. Liu, V. Gorovoy, S. Mukherjee, V. Kleiner, E. Koren, and E. Hasman,

[4] *Valley-Addressable Monolayer Lasing through Spin-Controlled Berry Phase Photonic Cavities*, Science **381**, 1429 (2023).

[5] A. Lininger, G. Palermo, A. Guglielmelli, G. Nicoletta, M. Goel, M. Hinczewski, and G. Strangi, *Chirality in Light–Matter Interaction*, Advanced Materials **35**, 2107325 (2023).

[6] Y. Chen et al., *Observation of Intrinsic Chiral Bound States in the Continuum*, Nature **613**, 7944 (2023).

[7] Y. Luo, C. Chi, M. Jiang, R. Li, S. Zu, Y. Li, and Z. Fang, *Plasmonic Chiral Nanostructures: Chiroptical Effects and Applications*, Advanced Optical Materials **5**, 1700040 (2017).

[8] S. Yang, Z. Liu, H. Yang, A. Jin, S. Zhang, J. Li, and C. Gu, *Intrinsic Chirality and Multispectral Spin-Selective Transmission in Folded Eta-Shaped Metamaterials*, Advanced Optical Materials **8**, 1901448 (2020).

[9] L. Kühner, F. J. Wendisch, A. A. Antonov, J. Bürger, L. Hüttenhofer, L. de S. Menezes, S. A. Maier, M. V. Gorkunov, Y. Kivshar, and A. Tittl, *Unlocking the Out-of-Plane Dimension for Photonic Bound States in the Continuum to Achieve Maximum Optical Chirality*, Light Sci Appl **12**, 1 (2023).

[10] M. V. Gorkunov, A. A. Antonov, and Y. S. Kivshar, *Metasurfaces with Maximum Chirality Empowered by Bound States in the Continuum*, Phys. Rev. Lett. **125**, 093903 (2020).

[11] M. Qiu, L. Zhang, Z. Tang, W. Jin, C.-W. Qiu, and D. Y. Lei, *3D Metaphotonic Nanostructures with Intrinsic Chirality*, Advanced Functional Materials **28**, 1803147 (2018).

[12] T. Shi et al., *Planar Chiral Metasurfaces with Maximal and Tunable Chiroptical Response Driven by Bound States in the Continuum*, Nat Commun **13**, 1 (2022).

[13] Z. Ma, Y. Li, Y. Li, Y. Gong, S. A. Maier, and M. Hong, *All-Dielectric Planar Chiral Metasurface with Gradient Geometric Phase*, Opt. Express, OE **26**, 6067 (2018).

[14] X. Zhang, Y. Liu, J. Han, Y. Kivshar, and Q. Song, *Chiral Emission from Resonant Metasurfaces*, Science **377**, 1215 (2022).

[15] M. S. Bin-Alam et al., *Ultra-High-Q Resonances in Plasmonic Metasurfaces*, Nat Commun **12**, 974 (2021).

[16] H. Qin, Z. Su, M. Liu, Y. Zeng, M.-C. Tang, M. Li, Y. Shi, W. Huang, C.-W. Qiu, and Q. Song, *Arbitrarily Polarized Bound States in the Continuum with Twisted Photonic Crystal Slabs*, Light Sci. Appl. **12**, 1 (2023).

[17] H. Qin, Y. Shi, Z. Su, G. Wei, Z. Wang, X. Cheng, A. Q. Liu, P. Genevet, and Q. Song, *Exploiting Extraordinary Topological Optical Forces at Bound States in the Continuum*, Sci. Adv. **8**, eade7556 (2022).

[18] X. Yin, J. Jin, M. Soljačić, C. Peng, and B. Zhen, *Observation of Topologically Enabled Unidirectional Guided Resonances*, Nature **580**, 7804 (2020).

[19] W. Liu, B. Wang, Y. Zhang, J. Wang, M. Zhao, F. Guan, X. Liu, L. Shi, and J. Zi, *Circularly Polarized States Spawning from Bound States in the Continuum*, Phys. Rev. Lett. **123**, 116104 (2019).

[20] K. Koshelev, S. Lepeshov, M. Liu, A. Bogdanov, and Y. Kivshar, *Asymmetric Metasurfaces with High-$Q$ Resonances Governed by Bound States in the Continuum*, Phys. Rev. Lett. **121**, 193903 (2018).

[21] K. Fang, Z. Yu, V. Liu, and S. Fan, *Ultracompact Nonreciprocal Optical Isolator Based on Guided Resonance in a Magneto-Optical Photonic Crystal Slab*, Opt. Lett. **36**, 4254 (2011).

[22] A. K. Zvezdin and V. A. Kotov, *Modern Magnetooptics and Magnetooptical Materials* (Institute of Physics Pub, Bristol : Philadelphia, Pa, 1997).

[23] G. Armelles, A. Cebollada, A. García-Martín, and M. U. González, *Magnetoplasmonics: Combining Magnetic and Plasmonic Functionalities*, Advanced Optical Materials **1**, 10 (2013).

[24] Z. Wang and S. Fan, *Magneto-Optical Defects in Two-Dimensional Photonic Crystals*, Appl. Phys. B

**81**, 369 (2005).

[25] S. Fan and Z. Wang, *An Ultra-Compact Circulator Using Two-Dimensional Magneto-Optical Photonic Crystals*, J. Magn. Soc. Jpn. **30**, 641 (2006).

[26] E. Yao, Z. Su, Y. Bi, Y. Wang, and L. Huang, *Tunable Quasi-Bound States in the Continuum in Magneto-Optical Metasurfaces*, J. Phys. D: Appl. Phys. **57**, 375104 (2024).

[27] D. O. Ignatyeva, D. Karki, A. A. Voronov, M. A. Kozhaev, D. M. Krichevsky, A. I. Chernov, M. Levy, and V. I. Belotelov, *All-Dielectric Magnetic Metasurface for Advanced Light Control in Dual Polarizations Combined with High-Q Resonances*, Nat Commun **11**, 5487 (2020).

[28] K. Yayoi, K. Tobinaga, Y. Kaneko, A. V. Baryshev, and M. Inoue, *Optical Waveguide Circulators Based on Two-Dimensional Magnetophotonic Crystals: Numerical Simulation for Structure Simplification and Experimental Verification*, Journal of Applied Physics **109**, 07B750 (2011).

[29] E. Jesenska, T. Yoshida, K. Shinozaki, T. Ishibashi, L. Beran, M. Zahradnik, R. Antos, M. Kučera, and M. Veis, *Optical and Magneto-Optical Properties of Bi Substituted Yttrium Iron Garnets Prepared by Metal Organic Decomposition*, Opt. Mater. Express **6**, 1986 (2016).

[30] W. Yang et al., *Observation of Optical Gyromagnetic Properties in a Magneto-Plasmonic Metamaterial*, Nat Commun **13**, 1719 (2022).

[31] F. Freire-Fernández, J. Cuerda, K. S. Daskalakis, S. Perumbilavil, J.-P. Martikainen, K. Arjas, P. Törmä, and S. van Dijken, *Magnetic on–off Switching of a Plasmonic Laser*, Nat. Photon. **16**, 1 (2022).

[32] N. Engheta, *Four-Dimensional Optics Using Time-Varying Metamaterials*, Science **379**, 1190 (2023).

[33] J. Ni, C. Huang, L.-M. Zhou, M. Gu, Q. Song, Y. Kivshar, and C.-W. Qiu, *Multidimensional Phase Singularities in Nanophotonics*, Science **374**, eabj0039 (2021).

[34] D. Kim, A. Baucour, Y.-S. Choi, J. Shin, and M.-K. Seo, *Spontaneous Generation and Active Manipulation of Real-Space Optical Vortices*, Nature **611**, 48 (2022).

[35] X. Zhao, J. Wang, W. Liu, Z. Che, X. Wang, C. T. Chan, L. Shi, and J. Zi, *Spin-Orbit-Locking Chiral Bound States in the Continuum*, Phys. Rev. Lett. **133**, 036201 (2024).

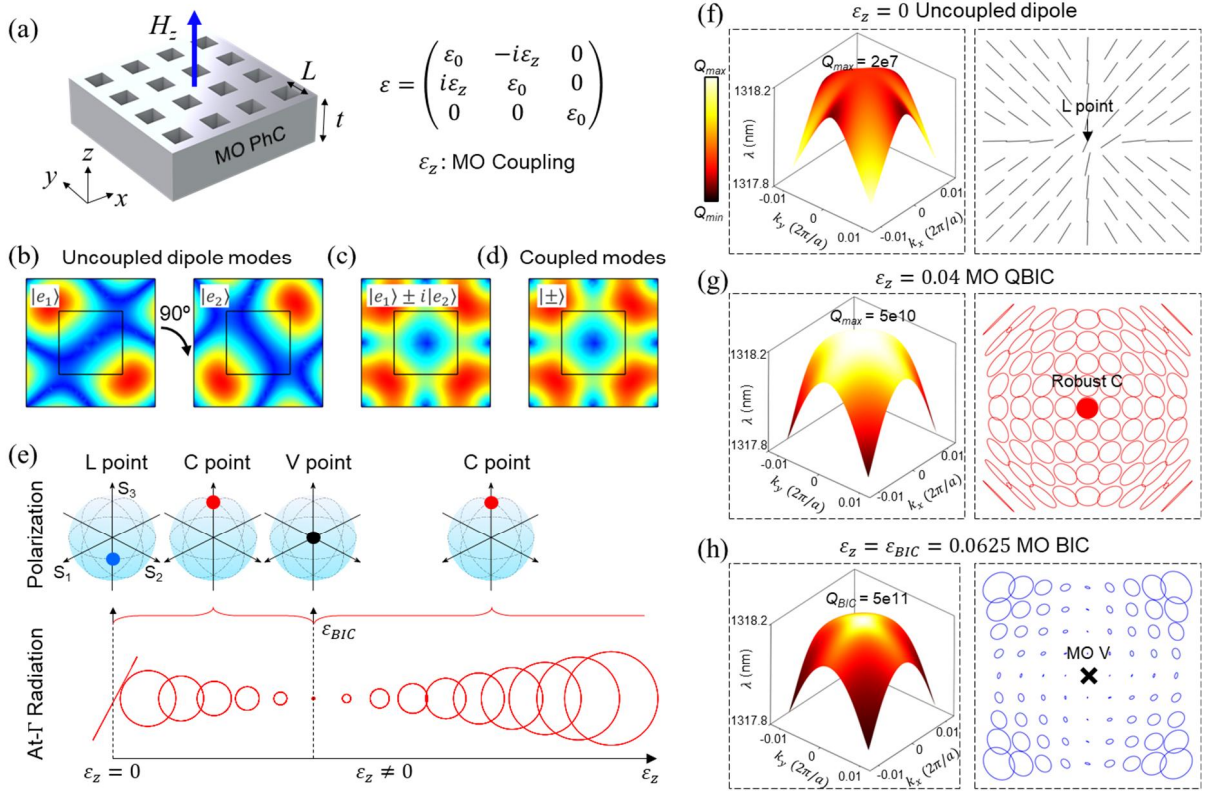

**Figure 1. Magneto-optical (MO) bound states in the continuum.** (a) Schematic of the MO PhC with squared air holes. The PhC is composed of MO material characterized by a permittivity tensor $\varepsilon$ with non-zero off-diagonal terms $\varepsilon_z$ induced by an external magnetic field along $z$-direction, facilitating MO coupling. The PhC has a periodicity of 540 nm, thickness $t = 665$ nm, and side length $L = 270$ nm for squared air holes. (b) Field profiles of two uncoupled dipole modes $|e_{1,2}\rangle$ are orthogonal with 90° rotational symmetry at $\varepsilon_z = 0$. (c) Field profile superposition of $|e_1\rangle \pm i|e_2\rangle$. (d) Field profile of simulated eigenmodes $|\pm\rangle$ at $\varepsilon_z \neq 0$. (e) Evolution of intrinsic eigenpolarizations on Poincaré sphere and corresponding radiation (polarization states and intensity) at $\Gamma$ point when varying $\varepsilon_z$ for high-Q $|-\rangle$ eigenmode. L and C points denote the radiation field with linear and circular polarization, respectively. MO BIC is achieved at $\varepsilon_{BIC}$ represented as V point. (f-h) Q factor and band structure (left panel), and polarization states (right panel) in momentum space of $|-\rangle$ eigenmode with (f) $\varepsilon_z = 0$, (g) $\varepsilon_z = 0.04$, and (h) $\varepsilon_z = 0.0625$, respectively.

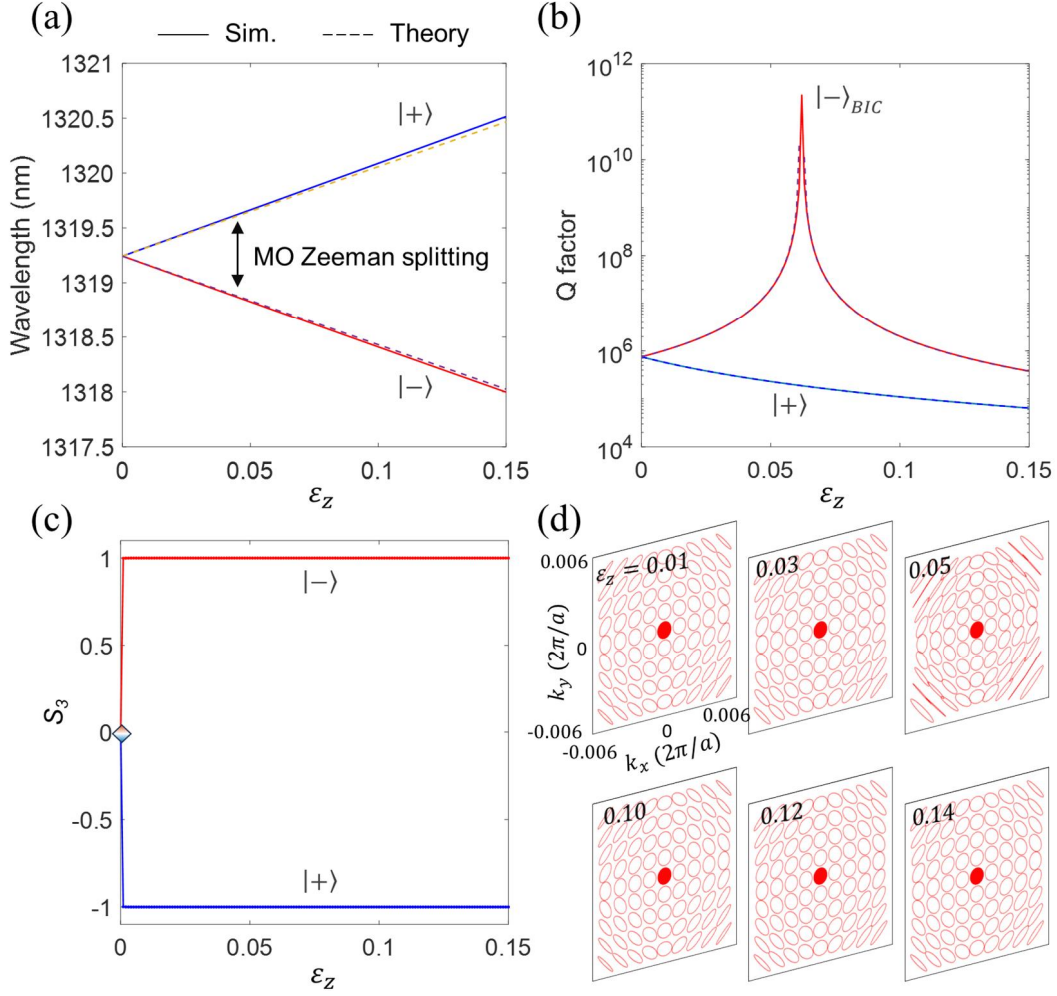

**Figure 2. Theoretical analysis on the working mechanism of MO BIC.** (a) The two modes degenerate at Dirac point when $\varepsilon_z = 0$. MO coupling induces Zeeman splitting, resulting in two circular eigenmodes $|\pm\rangle$ with non-zero $\varepsilon_z$. (b) Evolution of Q factor of two eigenmodes. MO BIC is realized with infinite Q factor at $\varepsilon_z = \varepsilon_{BIC} = 0.0625$ for $|-\rangle$ eigenmodes. Simulated (a) frequency and (b) loss of the two eigenmodes (solid lines) are consistent with the corresponding real and imaginary parts of the eigenvalues obtained from theory (dashed lines). (c) Stokes parameter $S_3$ of the two eigenmodes at Γ point maintains $\pm 1$ for $\varepsilon_z \neq 0$, implying stable circular polarization. (d) The polarization ellipses in momentum space for various $\varepsilon_z$ from 0.01 to 0.14, showing constant C points at Γ point.

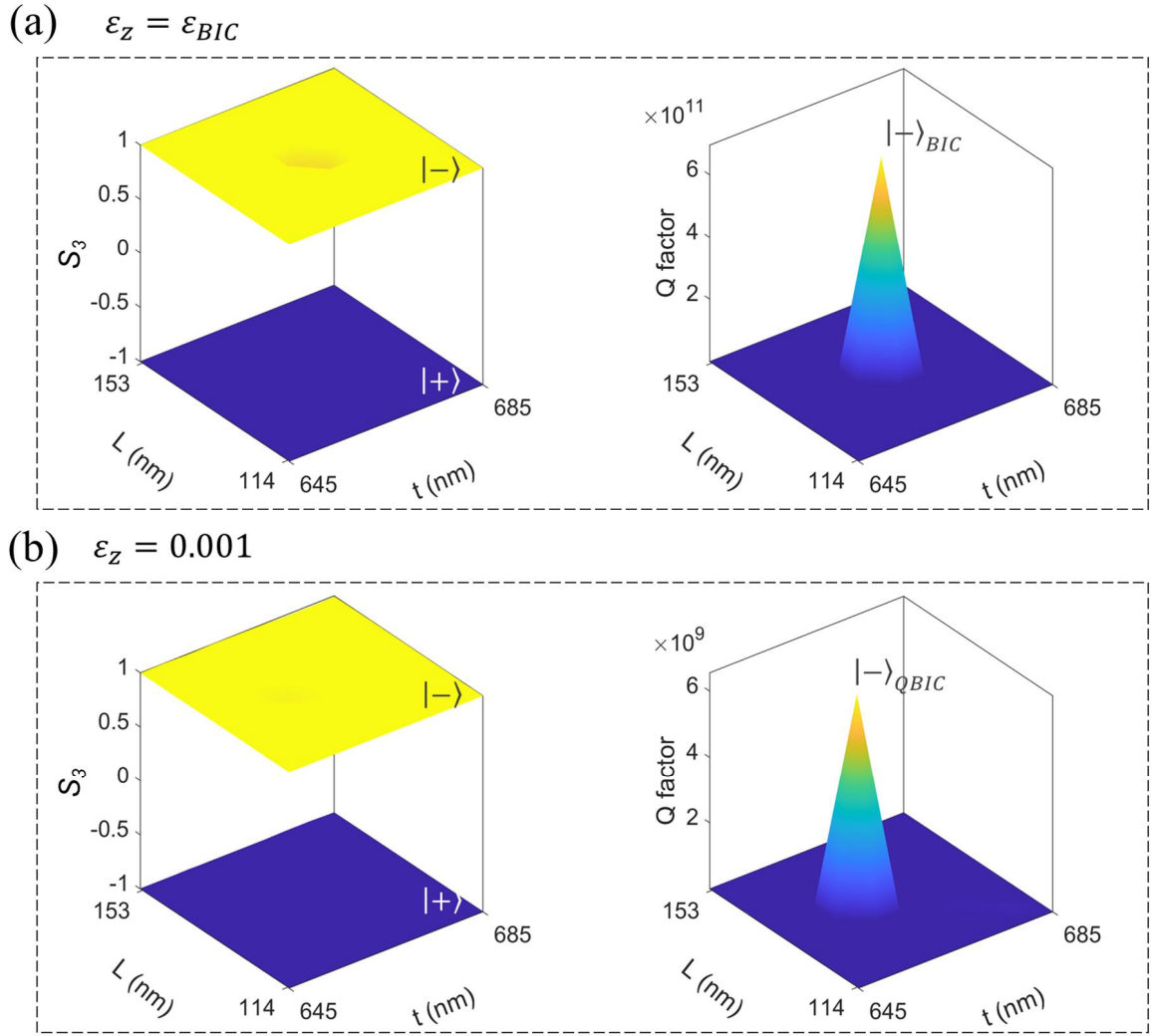

**Figure 3. Preservation of C points against structural geometry variations.** Far-field polarization state Stokes parameter $S_3$ and Q factor of MO eigenmode $|\pm\rangle$ in the parameter space of air hole side length and PhC thickness with (a) $\varepsilon_z = \varepsilon_{BIC}$ and (b) considerably small $\varepsilon_z = 0.001$. $S_3$ always maintains 1 for $|-\rangle$ and -1 for $|+\rangle$ except at the singularity of MO BIC.

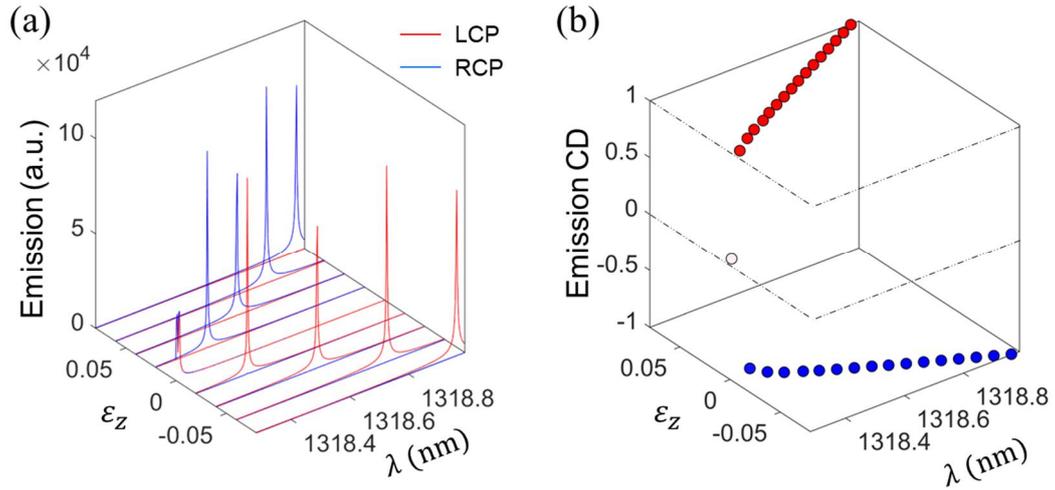

**Figure 4. Magnetically tunable chiral emission and circular dichroism (CD) near MO BIC.** (a) MO BIC exhibits tailorable intrinsic chiral emission of RCP when $\varepsilon_z > 0$ and LCP when $\varepsilon_z < 0$. (b) Emission CD maintains near 1 and -1 when $\varepsilon_z > 0$ and $\varepsilon_z < 0$, respectively.